\documentstyle[12pt]{article}
\advance \textheight by \topskip
\makeatletter
\def\eqnarray{\stepcounter{equation}\let\@currentlabel=\theequation
\global\@eqnswtrue
\global\@eqcnt\z@\tabskip\@centering\let\\=\@eqncr
$$\halign to \displaywidth\bgroup\@eqnsel\hskip\@centering
  $\displaystyle\tabskip\z@{##}$&\global\@eqcnt\@ne
  \hfil$\displaystyle{\hbox{}##\hbox{}}$\hfil
  &\global\@eqcnt\tw@ $\displaystyle\tabskip\z@
  {##}$\hfil\tabskip\@centering&\llap{##}\tabskip\z@\cr}
\@addtoreset{equation}{section}
  \def\theequation{\thesection.\arabic{equation}}
\makeatother
\def\Cal#1{{\cal #1}}
\def\u#1{\raisebox{-0.08ex}{$\displaystyle
  \mathop{u}^{\scriptscriptstyle #1}$}{}}
\def\tu#1{\raisebox{-0.08ex}{$\displaystyle
  \mathop{\widetilde u}^{\scriptscriptstyle #1}$}{}}
\def\p#1{\raisebox{-.35ex}{$\displaystyle
  \mathop{\varphi}^{\scriptscriptstyle #1}$}{}}

\begin{document}

\begin{titlepage}
\hbox to \hsize{\hfil hep-th/9305113}
\hbox to \hsize{\hfil IHEP 93--26}
\hbox to \hsize{\hfil February, 1993}
\vfill
\large \bf
\begin{center}
THE CANONICAL SYMMETRY AND \\
HAMILTONIAN FORMALISM. \\
I. CONSERVATION LAWS
\end{center}
\vskip 1cm
\normalsize
\begin{center}
{\bf A. N. Leznov and A. V. Razumov\footnote{E--mail:
razumov@mx.ihep.su}}\\
{\small Institute for High Energy Physics, 142284 Protvino, Moscow Region,
Russia}
\end{center}
\vskip 2.cm
\begin{abstract}
\noindent
The properties of the canonical symmetry of the nonlinear Schr\"odinger
equation are investigated. The densities of the local conservation laws
for this system are shown to change under the action of the canonical
symmetry by total space derivatives.
\end{abstract}
\vfill
\end{titlepage}

\section{Introduction}

The canonical symmetry of an integrable system is a special discrete
symmetry transformation having a number of remarkable properties
\cite{Lez91a}--\cite{LRa92}. In particular, as we have shown in
Ref.~\cite{LRa92}, this symmetry, considered as a transformation of the
phase space of the system, was a canonical transformation. In the same
paper, considering the case of the nonlinear Schr\"odinger equation, we
have established that the densities of the lowest conservation laws change
under the action of the canonical symmetry by total space derivatives.
Here we show that in this case all the conservation laws possess this
property. From this it follows, in particular, that the whole hierarchy of
the nonlinear Schr\"odinger equations is invariant with respect to the
action of the canonical symmetry.

Since, from our point of view, discrete transformations of the nonlinear
differential equations did not attract early the proper attention, we
found it useful to give a short review of the basic facts on such
transformations and their properties. This is done in sections 2 and 3.
Trying to make the paper self--contained but not too long, we decided to
give here the proofs only for those facts that could not be found in the
standard reference on the symmetry properties of the nonlinear
differential equations \cite{Olv86}.

In Ref.\ \cite{LRa92} to show that the canonical symmetry is a canonical
transformation we used the method of generating functions. At the
beginning of section 4 for the case of the nonlinear Schr\"odinger
equation we give another proof of this fact based on the representation of
the Poisson bracket with the help of the corresponding matrix differential
operator \cite{Olv86}. The rest of section 4 is devoted to the proof of
our main result concerning the behaviour of the densities of the
conservation laws under the canonical symmetry.

The summation over repeated indexes is implied.

\section{Hamiltonian Approach to Evolution Equations}

Here we recall the main facts on the Hamiltonian formulation of the
nonlinear evolution equations \cite{Olv86}. We restrict ourselves to the
case of two independent variables $t$, $x$ and $A$ dependent variables
$u^a$, $a = 1, \ldots, A$. An evolution equation is a system of equations
having the form
\begin{equation}
u^a_t = K^a(x, u, \u{(1)}, \ldots, \u{(K)}), \label{2.1}
\end{equation}
where for any $k$ $\u{(k)}$ denote a set formed by the $k$--th derivatives
$\u{(k)}^a$ of the dependent variables $u^a$ over $x$. It is convenient to
put $\u{(0)}^a \equiv u^a$.

Consider the space $\Cal A$ of the functions of the variables $x$, $u^a$,
and the derivatives of $u^a$ over $x$ up to some finite order. For a
general function of such type we use the following notation
\begin{equation}
f[u] = f(x, u, \u{(1)}, \ldots, \u{(K)}). \label{2.2}
\end{equation}
The operator of the total derivative over $x$ is defined as
\begin{equation}
D \equiv \frac{\partial}{\partial x} + \sum_{k=0}^\infty \u{(k+1)}^a
\frac{\partial}{\partial \u{(k)}^a}. \label{2.3}
\end{equation}
Actually, when the operator $D$ acts on an element of $\Cal A$ we have only
a finite number of members in series (\ref{2.3}).  For any two
functions $f$ and $g$ we have
\begin{equation}
D^m(fg) = \sum_{k=0}^m {m \choose k} D^{m-k}(f) D^k(g). \label{2.4}
\end{equation}

The operator
\begin{equation}
a[u] = \sum_{k=0}^{N(a)} a_k[u] D^k, \label{2.7}
\end{equation}
where $a_k \in \Cal A$, is called a differential operator of the order
$N(a)$\footnote{The operators, involving negative degrees of $D$, can also
be defined, but we will not use such operators in this paper.}. Using
Eq.~(\ref{2.4}), we can in an obvious way define the multiplication of
differential operators.  It can be shown that the multiplication of
differential operators is an associative operation.

The transposed (adjoint) operator of the differential operator $a$, given
by Eq.~(\ref{2.7}), is a differential operator $a^T$, defined as
\begin{equation}
a^T \equiv \sum_{k=0}^{N(a)} (-1)^k D^k a_k = \sum_{k=0}^{N(a)}
\sum_{m=k}^{N(a)} (-1)^m {m \choose n} D^{m-k} (a_m) D^k. \label{2.8}
\end{equation}
Here and below we use the following convention: if the operator $D$, or
some degree of the operator $D$, precedes some expression surrounded by
brackets, it acts only on the expression in the brackets.

A differential operator $a$ is called symmetric (self--adjoint) if $a^T =
a$, and it is called skew--symmetric (skew--adjoint) if $a^T = -a$.

A matrix with the matrix elements being differential operators is called a
matrix differential operator. Any matrix differential operator $A$
can be written as
\begin{equation}
A = \sum_{k=0}^{N(A)} A_k D^k, \label{2.9}
\end{equation}
where $A_k$ are matrices with the matrix elements being elements of $\Cal
A$. The transposed operator $A^T$ of the matrix differential operator $A$,
given by Eq.~(\ref{2.9}), is defined as
\begin{equation}
A^T = \sum_{k=0}^{N(A)} (-1)^k D^k A_k^T, \label{2.10}
\end{equation}
where $A_k^T$ is the transposed matrix of the matrix $A_k$.

Introduce on $\Cal A$ the operators of the variational derivatives over
$u^a$, defined as
\begin{equation}
\frac{\delta}{\delta u^a} \equiv \sum_{k=0}^\infty (-1)^k D^k
\frac{\partial}{\partial \u{(k)}^a}. \label{2.11}
\end{equation}
It can be shown that
\begin{equation}
\frac{\delta f}{\delta u^a} = 0, \qquad a = 1, \ldots, A, \label{2.12}
\end{equation}
if and only if there exists a function $g$, such that
\begin{equation}
f = Dg. \label{2.13}
\end{equation}
Hence, defining
\begin{equation}
\mbox{Ker } \delta/\delta u \equiv \left\{ f \in \Cal A \mid
\delta f/\delta u^a = 0,\; a = 1, \ldots, A \right\},
\label{2.14}
\end{equation}
we can write
\begin{equation}
\mbox{Ker } \delta/\delta u = D \Cal A. \label{2.15}
\end{equation}

Two functions $f$ and $g$ from $\Cal A$ are said to be equivalent if there
exists a function $h$ such that
\begin{equation}
f = g + Dh. \label{2.16}
\end{equation}
It is easy to prove that we have defined an equivalence relation. The
corresponding equivalence classes are called functionals. We denote the
equivalence class containing a function $f$ by $F$. In accordance with the
established tradition we write in this case $F = \displaystyle \int
f\,dx$. The space of all functionals is denoted by $\Cal F$. If a
functional $F$ is given we denote by $f$ an arbitrary representative from
the corresponding equivalence class and call $f$ the density of the
functional $F$.

A bilinear map $\{\cdot, \cdot \}: \Cal F \times \Cal F \to \Cal F$, is
said to be a Poisson bracket on $\Cal F$ if it satisfies the relations
\begin{eqnarray}
&\{F, G\} = - \{G, F\},& \label{2.17} \\
&\{F, \{G, H\}\} + \{G, \{H, F\}\} + \{H, \{F, G\}\} = 0& \label{2.18}
\end{eqnarray}
for any functionals $F$, $G$ and $H$. Equality (\ref{2.18}) is called the
Jacobi identity.

A matrix differential operator $J$ with the matrix elements $J^{ab}$, $a,b
= 1, \ldots, A$, is said to be a Hamiltonian operator if the equality
\begin{equation}
\{F, G\} = \int \frac{\delta f}{\delta u^a} J^{ab} \frac{\delta g}{\delta
u^b}\,dx \label{2.19}
\end{equation}
defines a Poisson bracket on $\Cal F$. It is clear that the right--hand
side of equality (\ref{2.19}) does not depend of the choice of $f$ and
$g$. The Poisson bracket (\ref{2.18}) can be shown to satisfy
Eq.~(\ref{2.17}) if and only if the operator $J$ is skew--symmetric. The
Jacobi identity imposes more complicated conditions on the operator $J$.
We do not discuss them here, and only note that if $J$ is a constant
matrix then the Jacobi identity is valid.

We say that evolution equation (\ref{2.1}) can be written in the
Hamiltonian form if there exists a Hamiltonian operator $J$ and a function
$h \in \Cal A$ such that
\begin{equation}
K^a = J^{ab} \frac{\delta h}{\delta u^b}. \label{2.20}
\end{equation}
If this is the case, we write the equations of motion in the form
\begin{equation}
u^a_t = J^{ab}[u] \frac{\delta h[u]}{\delta u^b} \label{2.21}
\end{equation}
and call equations (\ref{2.21}) the Hamilton equations. Actually Hamilton
equations (\ref{2.21}) are defined not by the function $h$ but by the
corresponding functional $H$, called the Hamiltonian of the system.

A conservation law of evolution equation (\ref{2.1}) is a relation
of the form
\begin{equation}
p_t = Df, \label{2.22}
\end{equation}
that should be valid in virtue of equation (\ref{2.1}). In Eq.~(\ref{2.22})
$p$ and $f$ are elements of $\Cal A$, which are called the density and the
flow of the conservation law, respectively. If we have conservation law
(\ref{2.22}) we say that the functional $P$ is a conserved quantity, or an
integral of motion.  For the case of Hamilton equations (\ref{2.21}), the
functional $P$, satisfying the equality\footnote{We allow $P$ to depend
explicitly on $t$.}
\begin{equation}
\frac{\partial P}{\partial t} + \{P, H\} = 0, \label{2.23}
\end{equation}
is a conserved quantity.

\section{Differential Transformations}

Consider a transformation of the space described by $x$, dependent
variables $u^a$, and their $x$--derivatives, that is determined by the
relations
\begin{equation}
\widetilde u^a = \varphi^a[u] = \varphi^a(x, u, \u{(1)}, \ldots,
\u{(K)}). \label{3.1}
\end{equation}
We call such a transformation a differential transformation.
If there exists a transformation
\begin{equation}
u{}^a = \widetilde \varphi^a[\widetilde u], \label{3.2}
\end{equation}
such that
\begin{equation}
\widetilde \varphi^a[\varphi [u]] = u^a, \label{3.3}
\end{equation}
then we have an invertible transformation. The transformation law for the
variables $\u{(k)}^a$ is defined by
\begin{equation}
\tu{(m)}^a = \p{(m)}^a[u] \equiv D^m \varphi^a [u]. \label{3.4}
\end{equation}

Define the operator $\varphi^*$ that acts on a function $f[u] = f(x, u,
\u{(1)}, \ldots, \u{(N)})$ according to the rule
\begin{equation}
\varphi^*f[u] \equiv f[\varphi[u]] = f(x, \varphi[u], \p{(1)}[u],
\ldots, \p{(N)}[u]). \label{3.5}
\end{equation}
It is not difficult to get convinced that for any $f \in \Cal A$ we have
\begin{equation}
D\varphi^*f = \varphi^*Df. \label{3.6}
\end{equation}

Let $\chi \in \Cal A^I$, i. e. $\chi$ is a set of functions $\chi^i \in
\Cal A$, $i = 1, \ldots, I$. The Fr\'echet derivative of $\chi$ is the $I
\times A$--matrix differential operator $\chi'$, defined as \cite{Olv86}
\begin{equation}
\chi'^i{}_a[u] v^a[u] \equiv \left. \frac{d}{d\epsilon} \chi^i[u +
\epsilon v[u]] \right|_{\epsilon = 0} \label{3.7}
\end{equation}
for any $v^a \in \Cal A$, $a = 1, \ldots, A$. From this definition we get
\begin{equation}
\chi'^i{}_a = \sum_{k=0}^\infty \frac{\partial \chi^i}{\partial
\u{(k)}^a} D^k. \label{3.8}
\end{equation}
In particular, for $f \in \Cal A$ we have
\begin{equation}
f'{}_a = \sum_{k=0}^\infty \frac{\partial f}{\partial \u{(k)}^a} D^k.
\label{3.9}
\end{equation}
It follows from this equality that
\begin{equation}
\frac{\delta f}{\delta u^a} = f'^T{}_a (1). \label{3.10}
\end{equation}

Let $\chi \in \Cal A^I$, denote the set of functions $\varphi^* \chi^i$,
$i = 1,\ldots, I$, by $\varphi^* \chi$. Let us show that
\begin{equation}
(\varphi^*\chi)'[u] = \chi'[\varphi[u]] \varphi'[u]. \label{3.11}
\end{equation}
Using Eq.~(\ref{3.8}) we get
\begin{equation}
(\varphi^*\chi)'^i{}_a = \sum_{n,m=0}^\infty \varphi^* \left(\frac{\partial
\chi^i}{\partial \u{(m)}^b}\right) \frac{\partial \p{(m)}^b}{\partial
\u{(n)}^a} D^n. \label{3.12}
\end{equation}
It is not difficult to show that
\begin{eqnarray}
\frac{\partial}{\partial \u{(n)}^a} D^m &=& \sum_{k=0}^m {m \choose k}
D^{m-k} \frac{\partial}{\partial \u{(n-k)}^a}, \qquad n > m, \label{3.13}
\\
\frac{\partial}{\partial \u{(n)}^a} D^m &=& \sum_{k=0}^n {m \choose k}
D^{m-k} \frac{\partial}{\partial \u{(n-k)}^a}, \qquad n \le m.
\label{3.14}
\end{eqnarray}
{}From this equalities and Eq.~(\ref{3.4}) it follows that
\begin{equation}
(\varphi^*\chi)'^i{}_a = \sum_{m=0}^\infty \varphi^*\left(\frac{\partial
\chi^i}{\partial \u{(m)}^b}\right) \left( \sum_{n=0}^m \sum_{k=0}^n +
\sum_{n=m+1}^\infty \sum_{k=0}^m \right) {m \choose k} D^{m-k} \left(
\frac{\partial \varphi^b}{\partial \u{(n-k)}^a} \right) D^n. \label{3.15}
\end{equation}
Now using the identity
\begin{equation}
\sum_{n=0}^m \sum_{k=0}^n = \sum_{k=0}^m \sum_{n=k}^m, \label{3.16}
\end{equation}
we can reduce Eq.~(\ref{3.15}) to
\begin{eqnarray}
(\varphi^*\chi)'^i{}_a &=& \sum_{n,m=0}^\infty \varphi^*\left(\frac{\partial
\chi^i}{\partial \u{(m)}^b}\right) \sum_{k=0}^m {m \choose k} D^{m-k}
\left( \frac{\partial \varphi^b}{\partial \u{(n)}^a} \right) D^{k+n}
\nonumber \\
&=& \sum_{m=0}^\infty \varphi^*\left(\frac{\partial
\chi^i}{\partial \u{(m)}^b}\right) D^m \sum_{n=0}^\infty \frac{\partial
\varphi^b}{\partial \u{(n)}^a} D^n. \label{3.17}
\end{eqnarray}
That was to be proved.

If differential transformation (\ref{3.1}) is invertible, then using
Eqs.~(\ref{3.3}) and (\ref{3.11}) we get
\begin{equation}
\widetilde \varphi'^a{}_c[\varphi[u]] \varphi'^c{}_b[u] = \delta^a_b.
\label{3.18}
\end{equation}

{}From Eq.~(\ref{3.11}) for any $f \in \Cal A$ we also have
\begin{equation}
(\varphi^*f)'{}_a [u] = f'{}_b[\varphi[u]] \varphi'^b{}_a[u]. \label{3.19}
\end{equation}
Taking into account Eq.~(\ref{3.10}), we get
\begin{equation}
\frac{\delta \varphi^*f}{\delta u^a} = \varphi'^T{}_a{}^b \varphi^*
\frac{\delta f}{\delta u^b}. \label{3.20}
\end{equation}

Consider now the behaviour of Hamiltonian equations (\ref{2.21}) under
differential transformations. Note first that from Eq.~(\ref{2.21}) it
follows that
\begin{equation}
\u{(n)}^a_t = D^n J^{ab}[u] \frac{\delta h[u]}{\delta u^b}. \label{3.21}
\end{equation}
For the transformed solution $\widetilde u^a$ we have
\begin{equation}
\widetilde u^a_t = \sum_{n=0}^\infty \frac{\partial \varphi^a[u]}{\partial
\u{(n)}^b} \u{(n)}^b_t = \varphi'^a{}_b[u] J^{bc}[u] \frac{\delta
h[u]}{\delta u^c}. \label{3.22}
\end{equation}
Let us suppose that we deal with an invertible transformation, then we can
define the differential operator $\widetilde J$ and the function
$\widetilde h$ by
\begin{eqnarray}
&\widetilde J[\varphi[u]] = \varphi'[u] J[u] \varphi'^T[u],& \\
&\widetilde h[\varphi[u]] = h[u].& \label{3.23}
\end{eqnarray}
In this case from Eq.~(\ref{3.20}) we get
\begin{equation}
\frac{\delta h[u]}{\delta u^a} = \frac{\delta \varphi^* \widetilde
h[u]}{\delta u^a} = \varphi'^T{}_a{}^b[u] \frac{\delta \widetilde
h[\varphi[u]]}{\delta u^b}. \label{3.24}
\end{equation}
Hence, equality (\ref{3.22}) can be written as
\begin{equation}
\widetilde u^a_t = \widetilde J^{ab}[\widetilde u] \frac{\delta \widetilde
h[\widetilde u]}{\partial u^b}. \label{3.25}
\end{equation}
Thus, if we have a solution of equations (\ref{2.21}), then after
transformation (\ref{3.1}) we get a solution of equations
(\ref{3.25}).

In the case when $\widetilde J = J$, i.~e. when
\begin{equation}
\varphi'[u] J[u] \varphi'^T[u] = J[\varphi[u]], \label{3.26}
\end{equation}
we call transformation (\ref{3.1}) a canonical transformation. If, in
addition,
$\widetilde h - h \in \mbox{Ker }\delta/\delta u$, i.e.
\begin{equation}
h[\varphi[u]]  - h[u] \in \mbox{Ker }\delta/\delta u, \label{3.27}
\end{equation}
then we call transformation (\ref{3.1}) a differential symmetry
transformation. In this case equations (\ref{3.25}) have the same form as
equations (\ref{2.21}).

\section{Nonlinear Schr\"odinger Equation}

In fact, here and Ref.~\cite{LRa92} we consider the following complex
extension of the nonlinear Schr\"odinger equation:
\begin{equation}
i \dot q + q'' - 2\epsilon r q^2 = 0, \qquad i \dot r - r'' + 2\epsilon
r^2 q = 0, \label{4.1}
\end{equation}
where $q$ and $r$ are arbitrary complex functions of the variables $x$ and
$t$, $\epsilon$ is the coupling constant. In Eq.~(\ref{4.1}) and below dot and
prime mean the partial derivative over $t$ and $x$, respectively.  The
canonical symmetry for this system has the form \cite{Lez91a,Lez92,SYa91}
\begin{equation}
\widetilde q = \frac{1}{\epsilon r}, \qquad \widetilde r = \epsilon r^2 q -
r'' + \frac{r'^2}{r}. \label{4.2}
\end{equation}
The inverse transformation is
\begin{equation}
q = \epsilon \widetilde q^2 \widetilde r - \widetilde q'' +
\frac{\widetilde q'^2}{\widetilde q},\qquad r = \frac{1}{\epsilon
\widetilde q}. \label{4.3}
\end{equation}

Denoting $u^1 \equiv q$, $u^2 \equiv r$, we can write equations (\ref{4.1})
in Hamiltonian form (\ref{2.21}), where
\begin{eqnarray}
&J = \left( \begin{array}{rr}
0 & -i  \\
i & 0 \end{array} \right),& \label{4.4} \\
&h[q,r] = r'q' + \epsilon r^2 q^2.& \label{4.5}
\end{eqnarray}

The differential operator $\varphi'$, corresponding to transformation
(\ref{4.2}), can be written as
\begin{equation}
\varphi' = \left(\begin{array}{cc}
\alpha & \beta \\
\gamma & \delta
\end{array}\right), \label{4.6}
\end{equation}
where
\begin{eqnarray}
&\alpha = 0, \qquad \beta = -\frac{1}{\epsilon r^2}, \qquad \gamma = \epsilon
r^2,& \label{4.7} \\
&\delta = - D^2 + 2 \frac{r'}{r} D + 2 \epsilon r q - \frac{r'^2}{r^2}. &
\label{4.8}
\end{eqnarray}
It is easy to show that in this case equality (\ref{3.26}) is valid, hence
transformation (\ref{4.2}) is a canonical transformation.

In Ref.~\cite{LRa92} we have shown that for $h$, given by Eq.~(\ref{4.6})
we have
\begin{equation}
h[\widetilde q, \widetilde r] - h[q,r] = D\left( irq - \frac{i}{2 \epsilon}
\frac{r'^2}{r^2} \right). \label{4.9}
\end{equation}
Thus, transformation (\ref{4.2}) is a symmetry transformation.

The densities of the conservation laws for the system, described by
Eq.~(\ref{4.1}), are given by the formula \cite{ZSh71,FTa87}
\begin{equation}
p_n[q,r] = r w_n[q,r], \qquad n = 1,2, \ldots, \label{3}
\end{equation}
where
\begin{equation}
w_1[q,r] = q, \label{4}
\end{equation}
and the quantities $w_n$ for $n > 1$ can be found from the recursive
relation
\begin{equation}
w_{n+1} = -i D (w_n) + \epsilon r \sum_{m=1}^{n-1} w_m w_{n-m}. \label{5}
\end{equation}
Multiplying Eq.~(\ref{5}) by $r$ and using Eq.~(\ref{3}), we get the
following recursive relation for the densities $p_n$:
\begin{equation}
p_{n+1} = -i D (p_n) + \frac{i r'}{r} p_n + \epsilon \sum_{m = 1}^{n-1} p_m
p_{n-m}, \label{6}
\end{equation}
with the initial condition
\begin{equation}
p_1[q,r] = rq. \label{7}
\end{equation}
It is convenient for our purposes to introduce a fictitious conservation law
\begin{equation}
p_0[q,r] = \frac{i r'}{\epsilon r} \label{8}
\end{equation}
and write Eq.~(\ref{6}) as
\begin{equation}
p_{n+1} = -i D(p_n) + \epsilon \sum_{m = 0}^{n-1} p_m p_{n-m}.  \label{9}
\end{equation}

Consider now the behaviour of the densities $p_n$ under transformation
(\ref{4.2}). Denote by $f_n$ the difference between the transformed density
and the initial one, i. e.
\begin{equation}
f_n[q, r] = p_n[\widetilde q, \widetilde r] - p_n[q, r]. \label{10}
\end{equation}
For $n = 0, 1$ we have
\begin{equation}
f_0 = \frac{i}{\epsilon} \left(\frac{iD^2 (p_0) + D(p_1)}{iD(p_0) +
p_1}\right) = \frac{i}{\epsilon}D (\ln (iD(p_0) + p_1)), \quad f_1 =
iD(p_0). \label{11}
\end{equation}
{}From Eq.~(\ref{9}) we get for $n > 1$ the following recursive relation
\begin{equation}
f_{n+1} = -iD(f_n) + \sum_{m = 0}^{n-1}(f_m p_{n-m} + p_m f_{n-m} + f_{m}
f_{n-m}). \label{12}
\end{equation}
Using this relation, we find subsequently
\begin{eqnarray}
f_2 &=& iD\left(p_1 + \frac{\epsilon}{2} p_0^2\right), \label{13} \\
f_3 &=& iD\left(p_2 + \epsilon p_0 p_1 + \frac{\epsilon^2}{3}
p_0^3\right), \label{14} \\
f_4 &=& iD\left(p_3 + \epsilon p_0 p_2 + \frac{\epsilon}{2} p_1^2 +
\epsilon^2 p_0^2 p_1 + \frac{\epsilon^3}{4} p_0^3\right). \label{15}
\end{eqnarray}
Thus, it is natural to suppose that for $n \ge 1$
\begin{equation}
f_n = D(d_n), \label{16}
\end{equation}
where the quantities $d_n$ are given by
\begin{equation}
d_n = i \sum_{m=1}^n \left(\frac{\epsilon^{m-1}}{m} \sum_{i_1 + \cdots +
i_m + m = n} p_{i_1} \ldots p_{i_m} \right). \label{17}
\end{equation}
Let us prove this statement.

{}From Eqs.~(\ref{16}) and (\ref{17}) it follows that
\begin{equation}
f_n = i \sum_{m=1}^n \left(\epsilon^{m-1} \sum_{i_1 + \cdots + i_m + m = n}
D(p_{i_1}) p_{i_2} \ldots p_{i_m}\right). \label{18}
\end{equation}
It is not difficult to show that
\begin{equation}
f_{n+1} = iD(p_n) + \epsilon \sum_{m=0}^{n-1} p_m f_{n-m}. \label{19}
\end{equation}
Hence, it is enough to show that the equality
\begin{equation}
i D(p_n + f_n) = \epsilon \sum_{m=0}^{n-1} f_m (p_{n-m} + f_{n-m})
\label{20}
\end{equation}
is true. From Eq.~(\ref{18}), using repeatedly Eq.~(\ref{9}), we get
\begin{equation}
p_n + f_n = (i D(p_0) + p_1) c_{n-1}, \label{21}
\end{equation}
where $c_0 = 1$, and for $n \ge 1$
\begin{equation}
c_n = \sum_{m=1}^{n} \left( \epsilon^m \sum_{i_1 + \cdots + i_m + m = n}
p_{i_1} \ldots p_{i_m}\right). \label{22}
\end{equation}
Taking into account Eq.~(\ref{22}) and the explicit expression for $f_0$, we
reduce Eq.~(\ref{20}) to the relation
\begin{equation}
D(c_{n-1}) = -i\epsilon \sum_{m=1}^{n-1} f_m c_{n-1-m}. \label{23}
\end{equation}
The validity of Eq.~(\ref{23}) follows directly from Eqs.~(\ref{18}) and
(\ref{22}).

Thus, we have shown that in the case of the nonlinear Schr\"odinger
equation the densities of the local conservation laws change under
the action of the canonical symmetry by total space derivatives. We call
such densities quasi--invariants of the differential transformation under
consideration. A direct consequence of the fact, we have proved, is the
invariance of the higher nonlinear Schr\"odinger equations under the action
of the transformations, given by Eq.~(\ref{4.2}).

\section{Conclusion}

It is likely that only the densities of the local conservation laws of the
nonlinear Schr\"odinger equation are quasi-invariant under the action of
transformations (\ref{4.2}). If it is the case, then it is interesting to
find the conditions under which the invariants of some discrete
differential transformation, being a canonical transformation, are
involutive with respect to the Poisson bracket.

We see, that the condition of the quasi--invariance under the canonical
symmetry can be considered as a characteristic of the densities of the
local conservation laws. Note that this fact does not give us a
constructive method to find such  densities. We can use for this the
recursion operator, which can be constructed with the help of a relevant
Hamiltonian pair \cite{Olv86}, corresponding to the considered system of
nonlinear equations. In the next paper we show, that the notion of the
canonical symmetry is useful in looking for such operators.

\end{document}